\documentclass[pre,aps,twocolumn,superscriptaddress]{revtex4-1}
\pdfoutput=1
\usepackage{amssymb}
\usepackage{amsmath}
\usepackage{subfigure}
\usepackage{graphicx}
\usepackage{textcomp}
\usepackage{url}
\usepackage{color}
\usepackage{cases}
\usepackage[normalem]{ulem}
\usepackage{float}

\usepackage[colorlinks=true, urlcolor=blue, anchorcolor=blue, citecolor=blue,filecolor=blue,linkcolor=blue,menucolor=blue
]{hyperref}

\usepackage[titletoc,title]{appendix}

\usepackage{color}

\begin{document}
\title{Observing symmetry-broken optimal paths of stationary Kardar-Parisi-Zhang interface via a large-deviation
  sampling of directed polymers in random media}

\author{Alexander K. Hartmann}
\email{a.hartmann@uni-oldenburg.de}
\affiliation{Institut f\"{u}r Physik, Universit{\aa}t Oldenburg - 26111 Oldenburg, Germany}
\author{Baruch Meerson}
\email{meerson@mail.huji.ac.il}
\affiliation{Racah Institute of Physics, Hebrew University of
Jerusalem, Jerusalem 91904, Israel}
\author{Pavel Sasorov}
\email{pavel.sasorov@gmail.com}
\affiliation{Institute of Physics CAS -- ELI Beamlines, 182 21 Prague, Czech Republic}
\affiliation{Keldysh Institute of Applied Mathematics, Moscow 125047, Russia}


\begin{abstract}

Consider the short-time probability distribution $\mathcal{P}(H,t)$ of the one-point interface height difference $h(x=0,\tau=t)-h(x=0,\tau=0)=H$ of the stationary interface $h(x,\tau)$ described by the Kardar-Parisi-Zhang equation.
It was previously shown that the optimal path --
the most probable history of the interface $h(x,\tau)$  which dominates the upper tail of $\mathcal{P}(H,t)$  -- is described by any of \emph{two} ramp-like structures of $h(x,\tau)$ traveling either to the left, or to the right. These two solutions emerge, at a critical value of $H$, via a spontaneous breaking of the mirror symmetry $x\leftrightarrow -x$ of the optimal path, and this symmetry breaking is responsible for a second-order dynamical phase transition in the system.   We simulate the interface configurations numerically
by employing a large-deviation Monte Carlo
sampling algorithm in conjunction with the mapping between the KPZ interface and the directed polymer in a random potential at high temperature.  This allows us
to observe the optimal paths, which determine each of the two tails of
$\mathcal{P}(H,t)$, down to probability densities as small as $10^{-500}$.
At short times we observe mirror-symmetry-broken  traveling optimal paths for the upper
tail, and a single mirror-symmetric path for the lower tail, in good quantitative agreement with analytical predictions. At long times, even at moderate values
of $H$, where
the optimal fluctuation method is \emph{not} supposed to apply,  we still observe two well-defined dominating paths. Each of them violates the mirror symmetry $x\leftrightarrow -x$ and is a mirror image of the other.

\end{abstract}

\maketitle

\section{Introduction}
\label{intro}

Large deviations of fluctuating quantities in stochastic macroscopic systems out of equilibrium continue to attract much interest from the statistical mechanics community.
Of special interest are situations where the large-deviation functions  exhibit singularities, see Refs.  \cite{Schuetz2001,Derrida2007,Hurtado2014,bertini2015,BK2015} for reviews. Among these
there are singularities which can be classified as dynamical phase transitions (DPTs) \cite{Bertini2005,BD2005,Bertini2006,Hurtado2011}, because the pertinent large-deviation functions involve time $t$, and $t\to \infty$.
Here we consider the celebrated Kardar-Parisi-Zhang (KPZ) equation \cite{KPZ}  in 1+1 dimension,
\begin{equation}
\partial_{\tau}h=
\nu\partial_{x}^{2}h+(\lambda/2)\left(\partial_{x}h\right)^{2}+\sqrt{D}\,\xi(x,\tau)\,. \label{KPZ1}
\end{equation}
It has been shown recently \cite{Janas2016,SKM2018} that the KPZ
equation also exhibits a DPT, but in the opposite limit of $t\to 0$. This DPT, and breaking of the mirror symmetry $x\leftrightarrow -x$ of the optimal paths of the interface, \textit{i.e.} the interface height histories $h(x,t)$ which contribute most to rare-event statistics which we will specify shortly
will be the focus of attention of this work.

The KPZ equation (\ref{KPZ1}) describes the stochastic dynamics of the  height
$h(x,\tau)$ at point $x$ and at time $\tau$ of an interface without overhangs, driven by a delta-correlated Gaussian noise $\sqrt{D}\xi(x,\tau)$ with zero mean:
\begin{eqnarray}
 \langle\xi(x,\tau)\rangle &=& 0\,, \nonumber\\
  \langle\xi(x_{1},\tau_{1})\xi(x_{2},\tau_{2})\rangle&=&\delta(x_{1}-x_{2})\delta(\tau_{1}-\tau_{2})\,.
  \label{deltacorr}
\end{eqnarray}
In Eq.~(\ref{KPZ1}) $\nu>0$ is the diffusion constant, and the non-linearity coefficient $\lambda$ can be set to be positive (or negative) without loss of generality.

The  fluctuating quantity of our interest here is the interface height difference,
$$h(x=0,\tau=t)-h(x=0,\tau=0)=H\,,
$$
at a specified point $x=0$ at time $t$ \cite{shift}. The complete statistics of the height difference  is encoded in the probability distribution
  $\mathcal{P}(H,t)$ which
has been the focus of multiple recent studies, see Refs. \cite{QS2015,S2015,Takeuchi} for reviews. Note that
  the distribution $\mathcal{P}(H,t)$ depends on the initial condition
  $h(x,\tau=0)$. Commonly studied are the sharp-wedge (also called droplet), flat and
  stationary initial conditions \cite{QS2015,S2015,Takeuchi}.

We will be mostly interested in the short-time regime, when the observation time $t$ is much smaller than the characteristic nonlinear time of the KPZ equation  $t_{\text{NL}}= \nu^5/(D^2\lambda^4)$.  
At short times typical fluctuations of the KPZ interface, which corresponds to the body of $\mathcal{P}(H,t)$,
are still Gaussian. However, the KPZ nonlinearity is already
fully manifest in the tails of $\mathcal{P}(H,t)$. The DPT \cite{Janas2016,SKM2018} occurs, in the short-time regime,  for the \emph{stationary} initial condition, where it is assumed that the interface has evolved for a long time prior to $\tau=0$. For the stationary interface, initial configurations $h(x,\tau=0)$ are sampled from a statistical ensemble of random realizations of a two-sided Brownian motion:
\begin{equation}\label{twosidedBM}
h(x,\tau=0) = \sqrt{\frac{D}{2\nu}}\, B(x)\,.
\end{equation}
Here $B(x)$ is the two-sided Wiener process with diffusion constant $1/2$, so that $\langle B^2(x)\rangle =|x|$ \cite{pinning}. The short-time probability distribution $\mathcal{P}(H,t)$ scales,  up to a pre-exponential factor,
as \cite{Janas2016,KLD2017}
\begin{equation}
-\ln\mathcal{P}\left(H,t\right)
\simeq\frac{\nu^{5/2}}{D\lambda^{2}\sqrt{t}}\,\,s\left(\frac{\left|\lambda\right|H}{\nu}\right)\,,
 \label{actiondgen}
\end{equation}
where the scale function $s(\dots)$ is given by the (rescaled) effective action along the optimal path of the interface, as we explain below.
The function
\begin{equation}\label{LDF}
S(H)= -\lim\limits_{t\to 0} \,\sqrt{t}\,\ln \mathcal{P}(H,t)=\frac{\nu^{5/2}}{D \lambda^2} \,\,s\left(\frac{|\lambda|H}{\nu}\right)
\end{equation}
is the short-time large deviation function.  Remarkably, $S(H)$ exhibits a second-order DPT (that is a jump in the second derivative $d^2S/dH^2$) at a critical value of $H$ such that $\lambda H_c/\nu=3.7063\dots$. The DPT originates from a spontaneous breaking of mirror symmetry of the optimal path which we will discuss shortly.  A rather complete analogy of this DPT with the classical mean-field second-order transition in  equilibrium was established in Ref. \cite{SKM2018}, where a proper phase order parameter was identified, and an effective Landau theory was developed.

Ref. \cite{Janas2016} also determined the large-$|H|$ asymptotics of $S(H)$
\begin{numcases}
{S(H)\simeq}
\frac{4\sqrt{2}\, \nu }{3D
|\lambda|^{1/2}}\,|H|^{3/2}, & $\lambda H\to +\infty$,
\label{Shightail}\\
\frac{4\sqrt{2|\lambda|}}{15 \pi
D}\,|H|^{5/2}, &$\lambda H\to - \infty$ ,
\label{Slowtail}
\end{numcases}
which correspond to the (stretched-exponential) upper and lower tails of
the distribution $\mathcal{P}(H,t)$.

In this paper
we report, to our knowledge first, observations of the \emph{optimal paths} of the KPZ interface, which determine the distribution tails (\ref{Shightail}) and (\ref{Slowtail}) for the stationary interface, in numerical simulations.
Here  is some background information. The results of Refs. \cite{Janas2016,SKM2018} were obtained by the optimal fluctuation method (OFM) (see also Ref. \cite{exactresults}). The OFM -- by now a standard asymptotic tool of theoretical physics -- relies on a saddle-point evaluation of the path integral of the KPZ equation (\ref{KPZ1}), conditioned on reaching
a specified height $H$ at short times (or reaching a sufficiently large height $H$ at \emph{any} time). This procedure brings about a
classical field theory, formulated as a conditional variational problem. The solution of this problem, obeying the proper initial and boundary conditions, is called the optimal path of the conditioned process. It represents a  time-dependent
deterministic  field which describes the most probable time history of the system, dominating the contribution of different paths to the statistics of interest. Once the optimal path is determined, the action along it (plus the ``cost" of the optimal initial condition, if this additional optimization is present) yields the scale function $s(\dots)$ which enters Eq.~(\ref{actiondgen}). The OFM has been extensively used for studying the single-point height statistics of the KPZ equation in different settings and dimensions and for different initial conditions \cite{KK2007,KK2008,KK2009,MKV,KMSparabola,MeersonSchmidt2017,SMS2018,MSV_3d,SmithMeerson2018,
MV2018,Asida2019,SMV2019,LinTsai2021,Lamarre2021}.  Until very recently, exact analytical solutions of the OFM equations for the KPZ interface were only possible when using additional small parameters, such as a very large
or very small $|H|$. But in Refs. \cite{KLD2021a, KLD2021b} the complete short-time large-deviation functions for $\mathcal{P}(H,t)$ have been
obtained analytically for the droplet, flat and stationary
initial conditions by masterfully exploiting exact integrability of the
OFM equations pointed out in Ref. \cite{Janas2016}.

Without losing generality, we will suppose that $\lambda>0$. Crucially, at $H>H_c$ the OFM predicts the existence of \emph{two} optimal paths of the interface, each leading to the same scaling function
$s(\dots)$ \cite{Janas2016,SKM2018}.  The two optimal solutions
bifurcate from a single solution at the critical point $H=H_c$
  via a spontaneous breaking of
  the mirror symmetry - the spatial reflection symmetry $x\leftrightarrow -x$, leading to the second-order DPT in  $s(\dots)$. Each of the two optimal solutions is a mirror-symmetric ``twin" of the other with respect to $x\leftrightarrow -x$. At large positive  $H$,
  each of the two solutions has the form of a ramp-like traveling structure of $h(x,\tau)$ or, equivalently, a shock-antishock pair of the interface slope $V(x,\tau)=\partial_x h(x,\tau)$.  For $-\infty<H<H_c$, that is below the transition,  the optimal path is unique and mirror symmetric with respect to $x=0$. 

Besides their role in the calculation of large-deviation functions, optimal paths provide a valuable insight into the physics of large deviations,  and this insight is inaccessible by other methods.  However, a direct
observation of optimal paths in conventional numerical simulations
(and of course in experiments) is often difficult due to the very low probability of the large deviations in question \cite{practical_guide2015}.
Fortunately, this difficulty can be overcome by using large-deviation algorithms, \textit{e.g.} based on importance sampling approaches, as was recently shown for the KPZ equation in Refs.  \cite{Hartmann2018,HMS1,HKLD}. In these works a mapping from the KPZ equation to the (discrete) directed polymer in a random potential at high temperature was employed, and Monte-Carlo simulations with an importance sampling algorithm were  performed. As a result, the tails of the height distribution of the KPZ interface  were measured, for different initial conditions and at different times, down to probabilities as small as $10^{-1000}$ \cite{Hartmann2018,HKLD}. In particular, Ref. \cite{HKLD} measured the whole short-time distribution of $\mathcal{P}(H,t)$  for the stationary interface and, in particular, verified the
  asymptotics (\ref{Shightail}) and (\ref{Slowtail}). In Ref. \cite{HMS1}
the interface configurations at $\tau=t/2$, corresponding to the height
distribution tails at $\tau=t$, were observed for the droplet initial condition. The observed configurations turned out to be in good
agreement with those predicted from the OFM \cite{KMSparabola}.

Here we employ the same simulation strategy as in Refs.~\cite{Hartmann2018,HMS1,HKLD}. Our primary goal is to observe the  optimal paths of the stationary KPZ interface which correspond to the distribution tails (\ref{Shightail}) and (\ref{Slowtail}). Apart from the short-time regime, where the OFM is well known to apply, we also consider a long-time regime where the mere existence of optimal paths has not been established.

Here is a layout of the remainder of the paper.  In Sec. \ref{predictions} we
extend and elaborate on theoretical predictions of Refs. \cite{Janas2016,SKM2018} for the optimal paths. In Sec. \ref{simstrategy} we present our simulation strategy. In Sec. \ref{shorttime} we compare the simulation results with theory at short times, where the OFM is expected to apply \cite{Janas2016,SKM2018}.  In Sec. \ref{longtime} we present the results of our simulations in a long-time  regime. This includes probing a range of moderate (not too large)
values of $H$, where one does not expect OFM to be applicable.
In Sec. \ref{summary} we briefly summarize and discuss the main results of this work.

\section{Optimal paths: theoretical predictions}
\label{predictions}

Here we present and extend the theoretical predictions of
Refs.~\cite{Janas2016,SKM2018}. To be in line with
the notation of Refs. \cite{Janas2016,SKM2018}, we will temporarily switch to the convention $\lambda<0$. The final results of this  section will be presented in a fully dimensional form with an arbitrary sign of $\lambda$.
Introduce rescaled variables $\tau/t \to \tau$,  $x/\sqrt{\nu t}\to x$ and
$|\lambda| h/\nu \to h$.
Then Eq.~(\ref{KPZ1}) becomes dimensionless:
\begin{equation}
\label{eq:KPZ_dimensionless}
\partial_{\tau}h=\partial_{x}^{2}h-\frac{1}{2}\left(\partial_{x}h\right)^{2}+\sqrt{\epsilon} \, \xi\left(x,\tau\right),
\end{equation}
where
\begin{equation}\label{epsilondef}
\epsilon=\frac{D\lambda^{2}\sqrt{t}}{\nu^{5/2}}=\left(\frac{t}{t_{\text{NL}}}\right)^{1/2}
\end{equation}
is the rescaled noise magnitude.
We choose the reference frame of the interface position so that
\begin{equation}
\label{eq:BC_0}
h\left(x=0,\tau=0\right)=0
\end{equation}
and condition the rescaled stochastic process $h(x,\tau)$ on the equality
\begin{equation}
\label{H}
h\left(x=0,\tau=1\right)=H\,,
\end{equation}
where $H$ is rescaled by $\nu / |\lambda|$. Formally, the OFM demands that
$\epsilon \to 0$, but later on we will also consider regimes of finite and even large $\epsilon$.   The saddle-point procedure brings about a minimization problem for an effective action functional $s[h(x,\tau)]$. For the stationary interface this functional includes two terms \cite{Janas2016}:
\begin{equation}\label{fullactionfunctional}
s[h(x,\tau)]=s_{\text{dyn}}[h(x,\tau)] + s_{\text{in}}[h(x,0)]\,,
\end{equation}
where
\begin{equation}
\label{eq:sdyn_def}
\!\!\!\!s_{\text{dyn}}[h(x,\tau)]\!=
\!\frac{1}{2}\int_{0}^{1}\!\!\!d\tau\int_{-\infty}^{\infty}\!\!\!dx\left[\partial_{\tau}h-\partial_{x}^{2}h+\frac{1}{2}\left(\partial_{x}h\right)^{2}\right]^{2}
\end{equation}
is the dynamical contribution, whereas
\begin{equation}
\label{cost}
s_{\text{in}}[h(x,0)]=\int_{-\infty}^{\infty}dx\left.\left(\partial_{x}h\right)^{2}\right|_{\tau=0}
\end{equation}
is the ``cost'' of the initial height profile (\ref{twosidedBM}) \citep{Janas2016}. The ensuing Euler-Lagrange equation can be recast
in the Hamiltonian form
\begin{eqnarray}
  \partial_{\tau} h &=& 
  \partial_{x}^2 h - \frac{1}{2} \left(\partial_x h\right)^2+\rho ,  \label{eqh}\\
  \partial_{\tau}\rho &=& 
  - \partial_{x}^2 \rho - \partial_x \left(\rho \partial_x h\right) ,\label{eqrho}
\end{eqnarray}
where $\rho(x,\tau)$, the optimal realization of the (rescaled) noise $\xi(x,\tau)$, plays the role of the ``momentum density" field, which is canonically conjugate to the ``coordinate density", which is the optimal path $h(x,\tau)$ itself.

The minimization over $h(x,\tau=0)$ yields an initial condition 
in the form of a relation between \textit{a priori} unknown $h(x,0)$ and $\rho(x,0)$ \citep{Janas2016}:
\begin{equation}
\label{eq:OFM_initial_condition}
\rho\left(x,\tau=0\right)+2\partial_{x}^{2}h\left(x,\tau=0\right)=\Lambda\delta\left(x\right).
\end{equation}
This condition is specific to the stationary interface.  The one-point condition~(\ref{H}) leads to the singular condition
\begin{equation}
\label{pT}
\rho\left(x,\tau=1\right)=\Lambda\,\delta\left(x\right),
\end{equation}
where $\Lambda$, which plays the role of a Lagrangian multiplier, is to be ultimately expressed through $H$.


Once the optimal path, including the optimal initial height profile, is determined, one can evaluate the rescaled total action (\ref{fullactionfunctional}), and obtain the scaling function $s=s(H)$ which enters Eqs.~(\ref{actiondgen}) and (\ref{LDF}).

In Refs. \cite{Janas2016} and \cite{SKM2018} asymptotic solutions of the OFM problem~(\ref{eqh})-(\ref{pT}) were obtained
in the two limits, corresponding to the upper and lower tails of $\mathcal{P}$ and described by Eqs.~(\ref{Shightail}) and  (\ref{Slowtail}), respectively. We now extend and elaborate on the pertinent optimal solutions for $h(x,t)$, to be compared with our simulations in Sec. \ref{shorttime}.

\subsection{Optimal path for the upper tail $\lambda H>0$}
\label{highpath}

\begin{figure} [ht]
\includegraphics[width=0.36\textwidth,clip=]{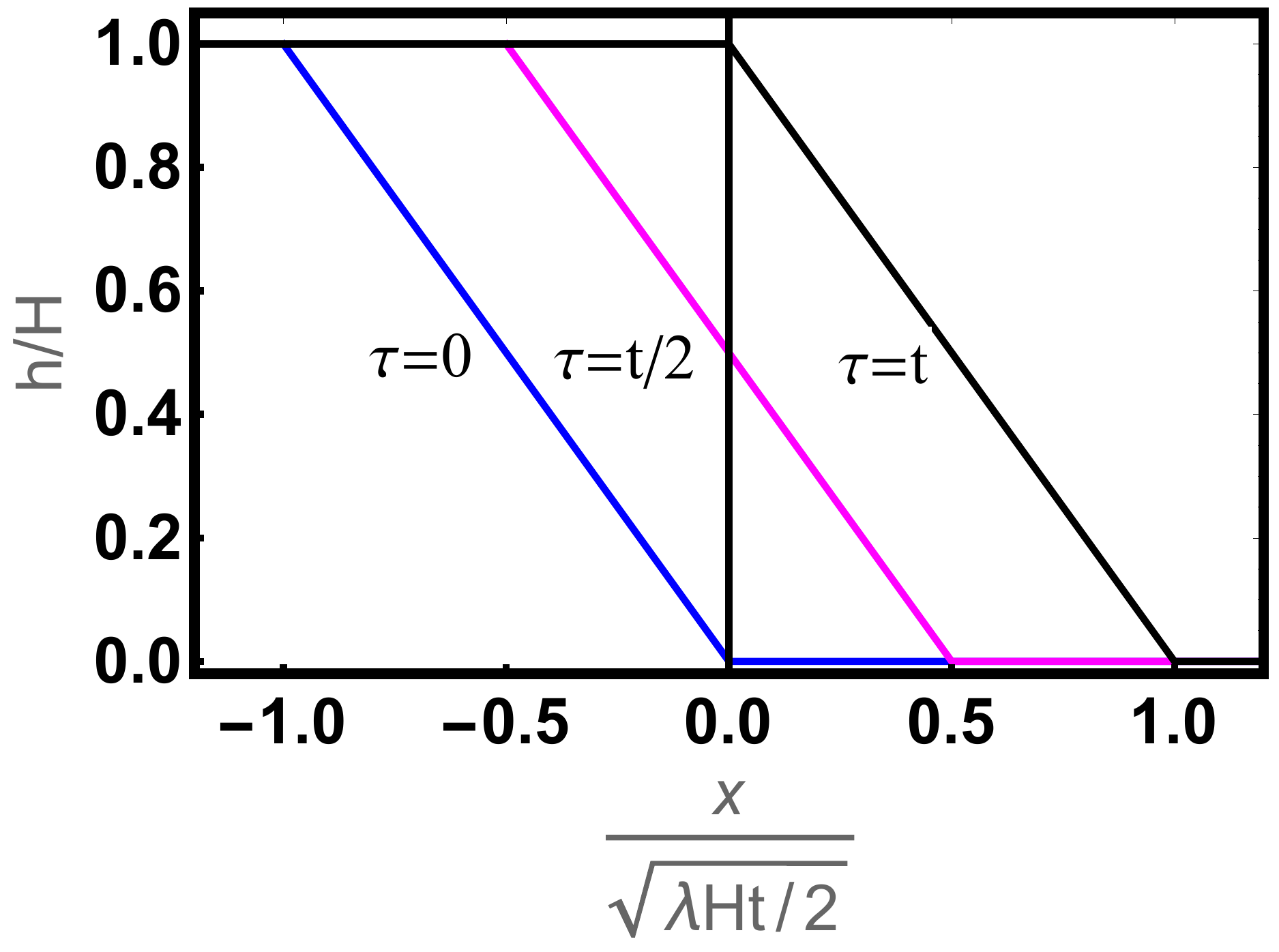}
\includegraphics[width=0.36\textwidth,clip=]{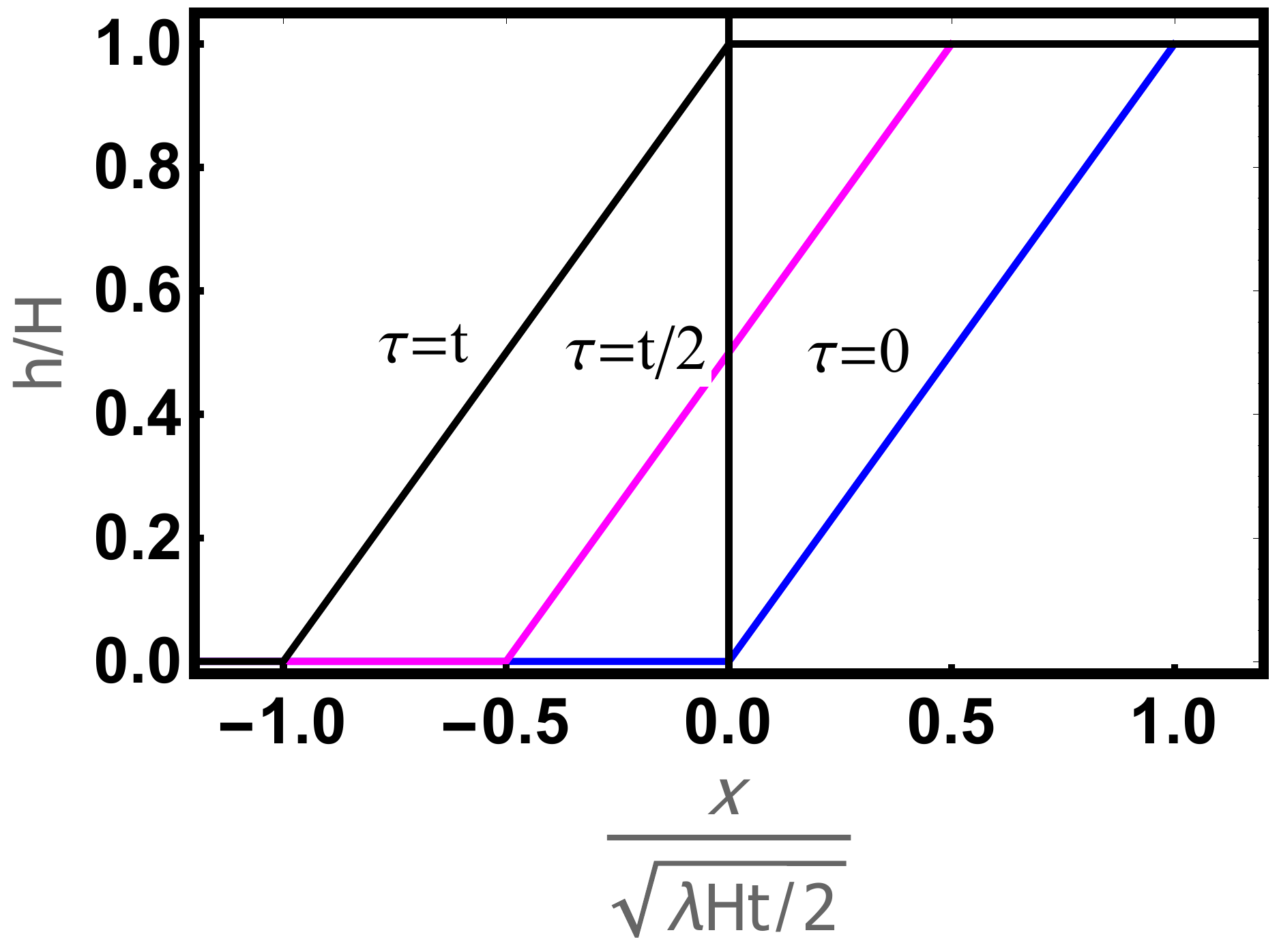}
\caption{Two mirror-symmetry broken optimal paths corresponding to the upper tail, $\lambda H>0$, of the height distribution $\mathcal{P}(H,t)$, see Eq.~(\ref{Shightail}). Shown is the leading-order prediction of Ref. \cite{Janas2016} for $h(x,\tau)$ as described by Eqs.~(\ref{ramp1})-(\ref{ramp3}) for the ramp traveling to the left (the right panel), and its mirror reflection with respect to $x=0$ for the ramp traveling to the right
(the left panel).}
\label{tworamps}
\end{figure}

Here we use the dimensional variables. At supercritical values of $H$ there are two mirror-symmetry broken optimal paths \cite{Janas2016}. In the $\lambda H \to +\infty$ tail, each of them represents a simple ramp-like structure for $h(x,\tau)$ [or,  equivalently, a shock-antishock pair of the interface slope $V(x,\tau)=\partial_x h(x,\tau)$], traveling  to the left or to the right. In terms of $h(x,\tau)$ the left-traveling ramp has the form \cite{Janas2016}:
\begin{numcases}
{\frac{h\left(x,\tau \right)}{H} \simeq}
1 ,& $x\geq \ell\,  \left(1-\mathfrak{t}\right)$,
\label{ramp1}\\
\frac{x}{\ell}+\mathfrak{t} , & $-\ell\, \mathfrak{t}\leq x\leq
\ell\, \left(1-\mathfrak{t}\right)$,
\label{ramp2}\\
0 , &$x\leq -\ell \,\mathfrak{t} $. \label{ramp3}
\end{numcases}
where $\mathfrak{t}=\tau/t$ and
\begin{equation}\label{ell}
\ell=\ell(H,t)=\left(\frac{|\lambda H| t}{2}\right)^{1/2}\,.
\end{equation}
 The right-traveling ramp solution can be obtained from Eqs.~(\ref{ramp1})-(\ref{ramp3}) by replacing $x$ by
$-x$.
Snapshots of the right- and left-traveling solutions for $h(x,\tau)$ are depicted in Fig.~\ref{tworamps}.

\subsection{Optimal path for the lower tail $\lambda H<0$}
\label{lowpath}
At $\lambda H\to -\infty$ the leading-order solution  is very different \cite{MKV,KMSparabola,Janas2016,SKM2018}. In the  ``pressure-dominated" region the solution describes an ideal hydrodynamic flow of an effective gas with density $\rho(x,\tau)$, velocity $V(x,\tau) = \partial_x h(x,\tau)$ and a negative pressure $-(1/2)\rho^2(x,\tau)$. There are also a region of
a zero-pressure hydrodynamic flow (also called the Hopf flow) \cite{SKM2018} and a trivial ``static region" where  $h(x,\tau)=H/2=\text{const}$.

\begin{figure} [ht]
\includegraphics[width=0.36\textwidth,clip=]{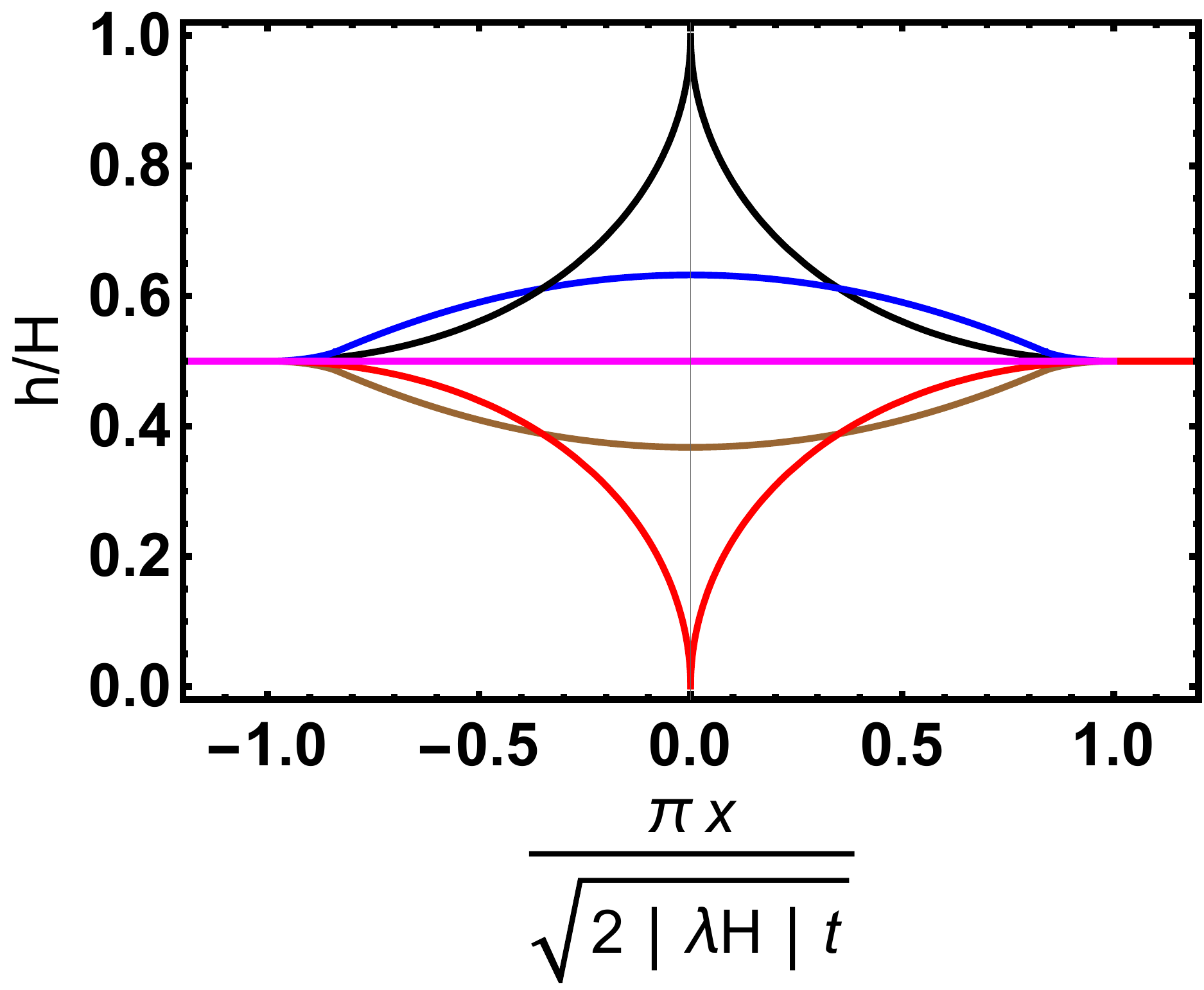}
\caption{The mirror-symmetric optimal paths corresponding to the lower tail, $\lambda H<0$, of the height distribution $\mathcal{P}(H,t)$, see Eq.~(\ref{Slowtail}). Shown are the leading-order theory prediction for $h(x,\tau)$ vs $x$ as described by Eqs.~(\ref{HD1})-(\ref{HD3}) at rescaled times
  $\mathfrak{t} = \tau/t = 0$ (black), $0.25$ (blue), $0.5$ (magenta), $0.75$ (brown),
and $1$ (red).}  
\label{HDtheoryfig}
\end{figure}

The results of Refs. \cite{MKV,KMSparabola,Janas2016,SKM2018} were presented in terms of $V(x,\tau)$ rather than $h(x,\tau)$. Here we calculated $h(x,\tau)$ by integrating $V(x,\tau)$ over $x$. After some algebra we obtained, in the three regions,
\begin{numcases}
{\!\!\!\frac{h\left(x,\tau \right)}{H} \simeq}
\mathfrak{h}_g\left(\frac{\pi x}{2\ell},\mathfrak{t}\right)
\!+\!\frac{1}{2},&\!\!\! $\!\frac{\pi |x|}{2}\!<\!\frac{\ell}{1+w(\mathfrak{t})^2} $,
\label{HD1}\\
\mathfrak{h}_v\left(\frac{\pi |x|}{2\ell},\mathfrak{t}\right)\!+\!\frac{1}{2}, & \!\!\! $\!\frac{\ell}{1+w(\mathfrak{t})^2}\!<\!\frac{\pi |x|}{2}\!<\!\ell$,
\label{HD2}\\
\frac{1}{2} , &\!\!\! $\! \frac{\pi |x|}{2}\!>\!\ell $. \label{HD3}
\end{numcases}
where $\ell$ is defined in Eq.~(\ref{ell}), and we have introduced three functions: $w$, $\mathfrak{h}_g$, and $\mathfrak{h}_v$. The function $w(\mathfrak{t})$ is the inverse of the function
\begin{equation}\label{h130}
\mathfrak{t}=\frac{1}{2}+\frac{w}{\pi(1+w^2)}+\frac{1}{\pi}\,\arctan w\,,
\end{equation}
the function $\mathfrak{h}_g(\chi,\mathfrak{t})$ is defined explicitly:
\begin{equation}\label{h095}
\mathfrak{h}_g(\chi,\mathfrak{t})=
\frac{1}{\pi}\arctan w\left(\mathfrak{t}\right)-\frac{1}{\pi}w\left(\mathfrak{t}\right)\left[1+w^2\left(\mathfrak{t}\right)\right]\chi^2\,,
\end{equation}
and the function $\mathfrak{h}_v(\chi,\mathfrak{t})$ is defined in a parametric form:
\begin{eqnarray}
   \mathfrak{h}_v(z,\mathfrak{t}) &\!=\!& u^2\left(\mathfrak{t}\!-\!\frac{1}{2}\right) \!+\!\frac{1}{\pi}\left[u+(u^2-1)\,
   \arctan u\right],\label{h100}\\
  z &\!=\!& \pi u\left(\mathfrak{t}\!-\!\frac{1}{2}\right) +1 \!+\!u \arctan u\,, \label{h110}
\end{eqnarray}
where the domain of $u$ in Eqs.~(\ref{h100})-(\ref{h110}) is defined by the double inequality
\begin{equation}\label{udomain}
-w(\mathfrak{t})\left(\mathfrak{t}-\frac{1}{2}\right)<u\left(\mathfrak{t}-\frac{1}{2}\right)<0\, .
\end{equation}
Overall, the optimal $h$-profiles (\ref{HD1})-(\ref{HD3}) as a function of
$x$ at different times $0\leq \tau\leq t$ are depicted in Fig. \ref{HDtheoryfig}. Notice that the optimal initial height profile has a plateau at $h=H/2$ at large $|x|$.
This property holds exactly for all subcritical values of $H$: $-\infty<\lambda H<|\lambda| H_c$, where the mirror symmetry of the optimal path is preserved.

\section{Simulations}
\label{simstrategy}

\subsection{Directed polymer mapping}
\label{map}

\begin{figure} [ht]
\includegraphics[width=0.36\textwidth,clip=]{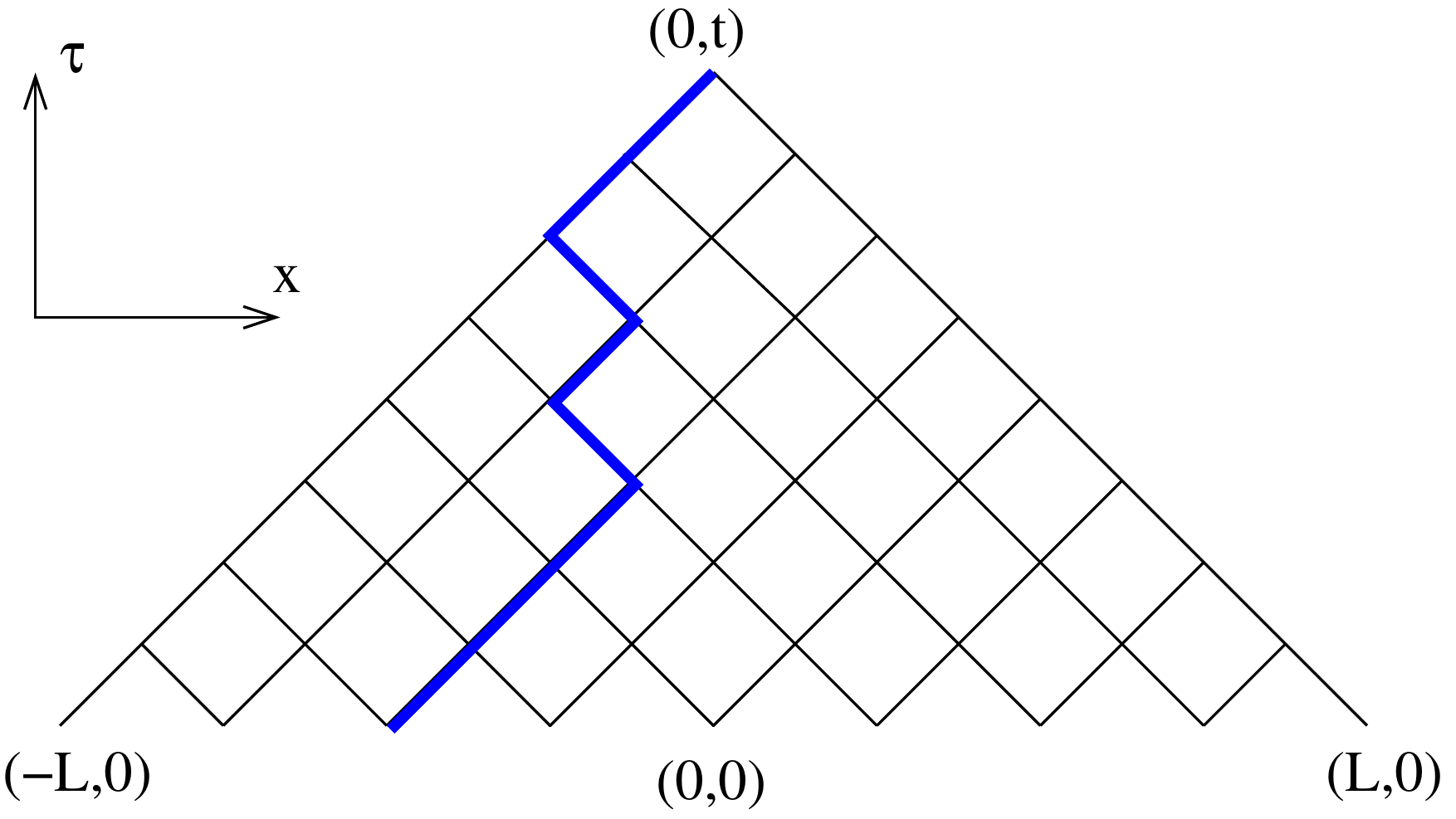}
\caption{A lattice realization of the directed polymer (see the text).
 $t=2L$.}
\label{dpolymer}
\end{figure}

Let us recall the lattice version of the mapping between the KPZ
height $h(x,t)$ and the free energy of a directed polymer in a
two-dimensional random potential  at high temperature $T$
\cite{Calabrese2010}. To follow the evolution of the height profiles in time,
it is convenient  to work with a
right-triangular domain, defined on a half square lattice with side length $L$
as shown in
Fig. \ref{dpolymer}. The lattice is indexed by points $(x,\tau)$ with
$\tau=0,1,\ldots, t$ ($t=2L)$ playing the role of a time and
$x=-x_{\max},-x_{\max}+1, \ldots,x_{\max}-1,
x_{\max}$ where $x_{\max}=(t-\tau)/2$.
 We consider all directed polymers which start at
a lattice point on the hypotenuse $(x,0)$ and end at the apex
$(0,2L)$.   The random value of the potential $V$ at each lattice
point is normally distributed with zero mean and unit
variance.  The partition function $Z(x,\tau)$
of a given realization of the potential
obeys the exact recursive equation \cite{Calabrese2010}
\begin{equation}\label{A004}
\!Z \left(x,\tau+1\right)\!=\! \left[Z\left(x-\frac{1}{2},\tau\right)+
  Z\left(x+\frac{1}{2},\tau\right)\right]\!
e^{-\frac{V\left(x,\tau+1\right)}{T}},
\end{equation}
where $\left\langle V\left(x,\tau\right)V\left(x^\prime,\tau^\prime\right)\right\rangle
=\delta_{xx^\prime}\delta_{\tau \tau^\prime}$, and $\delta_{xx^\prime}$
and $\delta_{\tau \tau^\prime}$ are Kronecker deltas.

Let us introduce $Z^*\left(x,\tau\right) = 2^{-\tau} Z\left(x,\tau\right)$. To obtain the mapping from the discrete Eq.~(\ref{A004}) to the continuous KPZ equation~(\ref{KPZ1}) we Taylor-expand all the discrete quantities in Eq.~(\ref{A004}), including $\exp(-V/T)$. For example,
\begin{equation}\label{taylor}
Z^*(x-\frac{1}{2},\tau)=Z^*(x,\tau)
-\frac{1}{2}\partial_{x}Z^*(x,\tau)
+\frac{1}{8}\partial_{x}^2Z^*(x,\tau)+\dots .
\end{equation}
To justify a truncation of this expansion
we must demand that $Z^*$ varies slowly on the scale of the lattice constant both in space and in time: $|Z^*(x,\tau+1)-Z^*(x,\tau)|\ll Z^*(x,\tau)$ and  $|Z^*(x+1,\tau)-Z^*(x,\tau)|\ll Z^*(x,\tau)$. This leads to the necessary conditions $L\gg 1$ and $|H|\ll L$. As we are studying large deviations, we consider $|H|\gg 1$. Finally, the slow variation of the  potential $V$ necessitates the condition $T\gg 1$. Overall, these necessary conditions  can be rewritten as a strong double inequality
\begin{equation}\label{mapconditions}
1\ll |H| \ll L\,.
\end{equation}
Under these conditions we can approximate the discrete equation~(\ref{A004}) by the continuous stochastic heat equation
\begin{equation}\label{A010}
\partial_{\tau} Z^*=\frac{1}{8}\partial_x^2Z^*-\frac{\xi}{T}Z^* \,.
\end{equation}
The (multiplicative) noise term $\xi$ is now continuous and delta-correlated in $x$ and $\tau$, and the mapping to the KPZ equation is given by the Cole-Hopf transformation
\begin{equation}
  h(x,\tau) = \ln [Z^*(x,\tau)/\langle Z^*(0,\tau) \rangle]\,,
  \end{equation}
where
\begin{equation}\label{parameters}
\lambda=1/4,\quad \nu =1/8, \quad  D=T^{-2} \quad  \text{and} \quad t= 2L.
\end{equation}
In particular, the height difference is given by
\begin{equation}
H=h(0,2L) = \ln [Z^*(0,2L)/\langle Z^*(0,2L) \rangle]\,. \label{eq:H}
  \end{equation}
The characteristic nonlinear time of the KPZ equation is
$t_{\text{NL}} =\nu^5/(D^2 \lambda^4)=T^4/2^7$, and $\epsilon = 16 \sqrt{L}/T^2$.

The Brownian initial condition   is specified by
\begin{equation}
  Z(x,0)  =  e^{-[V(x,0)+2 R(x)]/T}\,, \label{eq:Z:init}
\end{equation}
where $R(x)=\sum_0^{x} \eta(x)$ describes a random walk starting at the
origin $(0,0)$ and arriving at the site $(x,0)$ on the hypothenuse, where
$-L\le x\le L$. The random walk increments $\eta(x)$ are independent and normally distributed  random numbers with zero mean and variance $1$. At large values of $|x|$, the random walk $R(x)$ approaches the Brownian motion $B(x)$, and the coefficient $2/T$ in the exponent of Eq.~(\ref{eq:Z:init}) corresponds to the coefficient $D/(2\nu)$ in Eq.~(\ref{twosidedBM}).

\subsection{Importance sampling algorithm}
\label{ISA}

All measurable quantities, such as the height difference $H$,
depend on the set of calculated partition
function values $\{Z(x,\tau)\}$. Furthermore, the partition function
values depend through Eqs.~(\ref{A004}) and (\ref{eq:Z:init})
deterministically on the realization
$(V,\eta)$ of the quenched randomness,
with $V$ being the set of $(L+1)(2L+1)$ potential
values $\{V(-L,0),V(-L+1,0)$,\ldots,$V(L-1,0),V(L,0),V(-L+\frac 1 2,1),
V(-L+\frac 3 2,1)$, \ldots, $V(-\frac 1 2, 2L-1), V(\frac 1 2, 2L-1),
V(0,2L)\}$ and $\eta$ being the set of $2L+1$
random walk increments $\{\eta(-L),\eta(-L+1),$ \ldots,$\eta(L-1),\eta(L)\}$.
As we already mentioned, all these random values are independent and normally distributed with zero
mean and variance $1$.

A straightforward way to sample measurable quantities would be to generate
independent realizations $(V,\eta)$ of the disorder,
to calculate the partition function for each realization and obtain the desired quantity. In this
way estimates for expectation values can be obtained by averages,
possibly conditioned on certain values of a second observable, like $H$.
Estimates for distributions are obtained from histograms.

This procedure, however, would generate \emph{typical} values with respect to all
quantities, in particular typical values with respect to $H$. In the
present study we are rather interested in investigating the behavior
of the system
for very rare values of $H$. Thus,
we have to be able to address
the underlying distribution $\mathcal{P}(H)$
over a large part of the support, in particular  the low-probability tails.

To achieve this, we use a \emph{biased} distribution \cite{hammersley1956} of the randomness
by modifying the original quenched distribution
weight \cite{align2002}, which is a product of independent Gaussians,
by an additional exponential Boltzmann factor $e^{-H/\Theta}$, where $\Theta$ is
an adjustable  temperature-like
parameter, allowing us to address different regions of values of $H$.
When $\Theta \to \pm \infty$, we restore the
original unbiased distribution. For $\Theta \to 0^+$, one will focus
the sampling on large negative values of $H$, while for
$\Theta\to 0^-$ the sampling will be in the region of large positive values of $H$.
The fundamental idea of this large-deviation approach is versatile, and it
has been applied to various problems, e.g., to
study large-deviation properties of random graphs \cite{largest-2011,diameter2018},
of random walks \cite{fBm_MC2013,convex_hull2015}
of energy grids \cite{power_flow2015,stability2019},
of biological sequence alignments \cite{align2002,align_long2007},
of nonequilibrium work distributions \cite{work_ising2014}
and of traffic models \cite{schreckenberg2019}.

Note that any obtained statistics will initially follow the biased weight, but this
can be easily corrected by reweighting with the inverse weight, \textit{i.e.}
$e^{+H/\Theta}$,
and proper normalization \cite{align2002}. This reweighting is important
if one is interested in actually measuring $\mathcal{P}(H)$. In the present case,
where we only measure some
other quantities, conditioned on certain, e.g., extreme values of $H$, the reweighting is unnecessary. Thus,
we only have to be able to sample with respect to selected values of $H$,
which is achieved by the bias.

In the present case, the biased sampling
of the disorder realizations $(V,\eta)$ cannot be performed directly. Therefore
we applied Markov-chain Monte Carlo (MC) simulations \cite{newman1999}, where a
configuration
of the Markov chain at  step $s$ is given by a disorder
realization $(V,\eta)^{(s)}$. As explained above,
the corresponding height depends deterministically on the disorder
realizations, \textit{i.e.} $H^{(s)}=H[(V,\eta)^{(s)}]$.
Following the Metropolis-Hastings algorithm,
in each step some of the entries of $(V,\eta)^{(s)}$ are redrawn according to
the underlying Gaussian distribution, leading to a trial configuration
$(V,\eta)'$ which corresponds to a value $H'$. This value has to be obtained
in each MC step by a full calculation of the partition function using
Eqs.~(\ref{A004}) and (\ref{eq:Z:init}) and finally evaluating $H'$
with Eq.~(\ref{eq:H}). To obtain a sampling according to the
biased distribution, the trial configuration
is accepted, \textrm{i.e.} $(V,\eta)^{(s+1)}=(V,\eta)'$, with the usual Metropolis
probability $p_{\rm acc}=\min\{1, e^{-\Delta H/\Theta} \}$
depending on the change $\Delta H=H'-H^{(s)}$.
Otherwise the trial configuration is rejected, \textrm{i.e.}
$(V,\eta)^{(s+1)}=(V,\eta)^{(s)}$.
As usual, one has to make the MC simulations long enough to achieve equilibration and avoid correlated configurations. For more details on the implementation of the large-deviation algorithm for the KPZ model, including the choice of the values for $\Theta$, the combination of the results obtained for different
values of $\Theta$, the use of the parallel tempering approach, and the parallelization using Message Passing Interface, see Refs.~\cite{Hartmann2018,HKLD}.

The simulations where performed for the directed polymer model
  with length $t=2L=128$ and two values of temperature: $T=2$ and $T=8$.
We performed a parallel tempering simulation for a total of 148
different values of the temperatures $\Theta$ for $T=2$. For $T=8$
we used 487 values of temperatures. We used
a high-performance computer which required a total of more than
94000 core hours. During the parallel simulation, we
monitored equilibration by following the convergence of the
statistics of $H$. The convergence takes different time scales depending on the values of $\Theta$. During the simulation we stored
for $T=2$ more than 1500 configurations  with corresponding
  values of $H$ in the range $[-7,80]$. For $T=8$
  we stored about 23000 configurations with values of $H$
  ranging in the interval $[-58,86]$.
The configurations were spread with respect to $H$ more or less equally over the sampling intervals, but with an emphasis on the extremes of the
intervals, because more values of the temperature-like parameter $\Theta$ are needed for equilibrium sampling in this range.

\section{Theory versus simulations: short times}
 \label{shorttime}


Here we present the simulation results and compare them with the theoretical predictions from the OFM for the set of parameters~(\ref{parameters}) with $T=8$ and $L=64$. For these parameters $\epsilon=2$ 
is not a small number but, as was found earlier~\cite{Hartmann2018,HMS1,HKLD}, it still corresponds to the short-time limit. We will start from the case of $H>0$, where breaking of
the mirror symmetry is expected to occur, and then proceed to the case of $H<0$ where we expect the mirror symmetry to be preserved.

\subsubsection{Large positive $H$}

\begin{figure} [ht]
\includegraphics[width=0.36\textwidth,clip=]{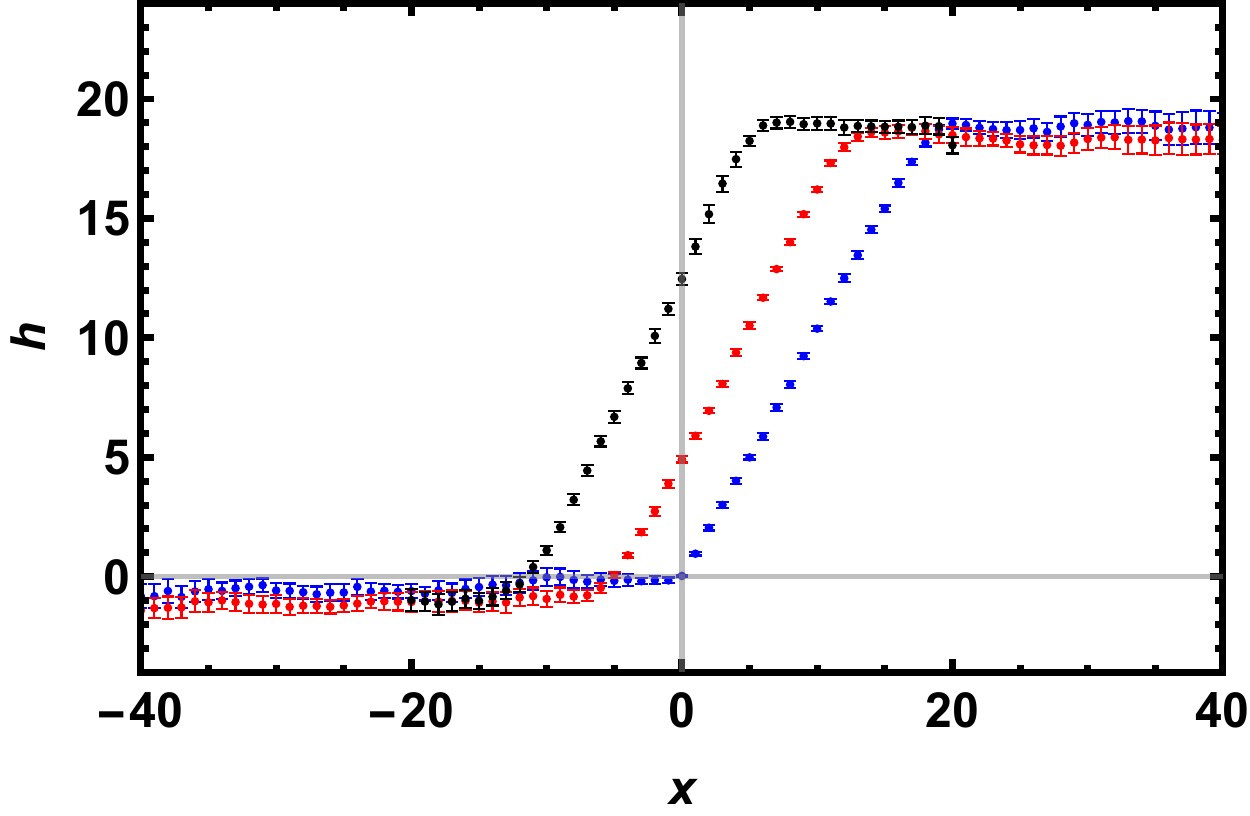}
\caption{The average over 9 simulated height profiles $h(x,\tau)$, conditioned on $H\simeq 20$ at rescaled times $\tau/t=0$ (blue), 0.35 (red), and 0.65 (black). The actual values of $H$ are $H\in [19.95, 20.1]$.  Eight of these height profiles are actually \emph{right}-travelling ramps flipped around $x=0$.  The error bars show the standard deviation. The simulation parameters $T=8$ and $L=64$ correspond to a short time regime.}
\label{high_tr}
\end{figure}

\begin{figure} [ht]
\includegraphics[width=0.36\textwidth,clip=]{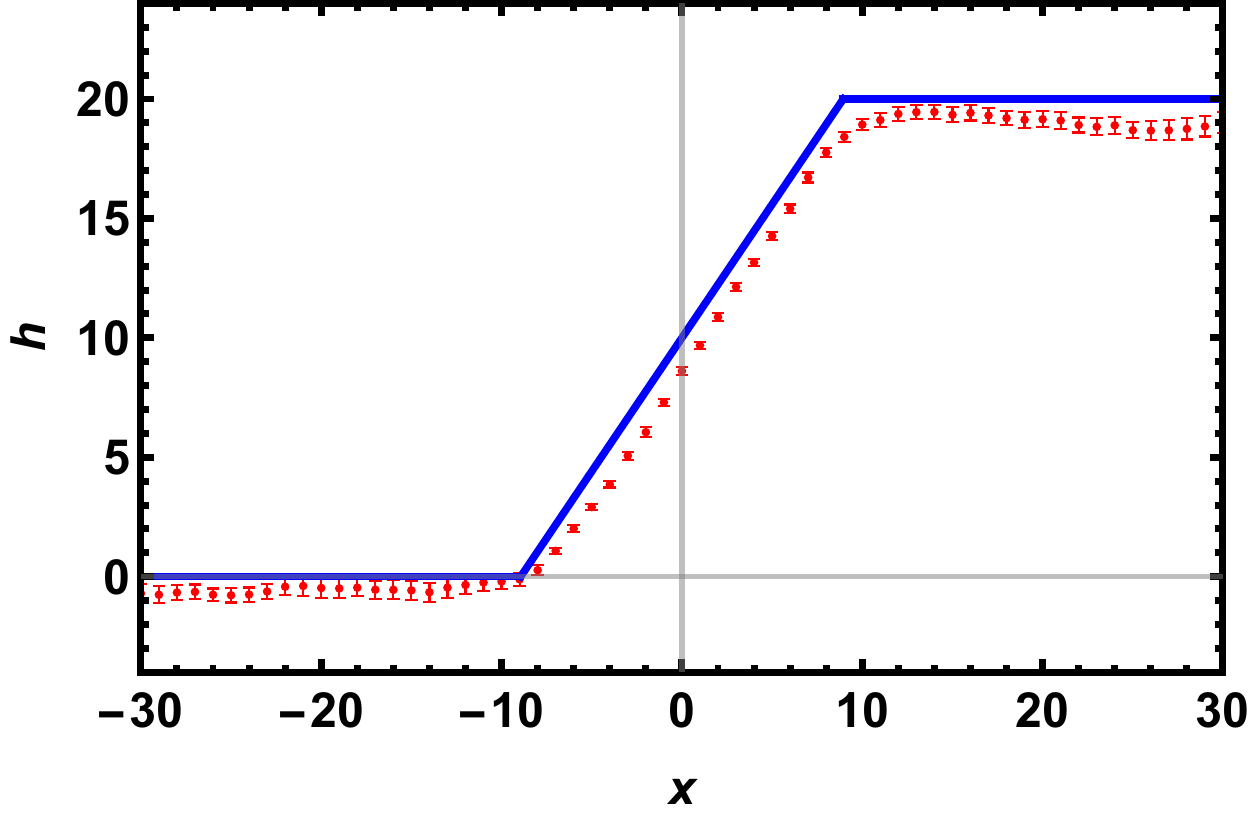}
\caption{A comparison,  at time $\tau=0.5 t$,  of the simulated average
height profile conditioned on $H\simeq 20$ (see Fig. \ref{high_tr}) with the asymptotic solution (\ref{ramp1})-(\ref{ramp3}) for the optimal path  \cite{Janas2016}.}
\label{high_com}
\end{figure}

In the range $10\le H\le 30$, we obtained 935
independent realizations $(V,\eta)$, each allowing for an analysis
of the trajectories of
$h(x,\tau)$ for all values $\tau \in[0,1]$. (Note that for standard sampling, realizations
exhibiting such values of $H$ would occur with their natural probabilities which are
$10^{-100}$ or smaller \cite{HKLD},
so we indeed needed to employ large-deviations algorithms to study
this range of $H$.)  We clearly observed the mirror
symmetry breaking in the form of left- and right-traveling ramps of $h(x,\tau)$.
To improve the statistics, we exploited the mutual reflection symmetry of the left- and right-traveling ramps with respect to each other: We flipped around $x=0$ the observed
$h(x,\tau)$ profiles of the right-travelling
ramp realizations  and processed them together with the
left-travelling-ramps.
The resulting Fig.~\ref{high_tr}  shows the
average over all the simulated  $h(x,\tau)$-profiles for $H \simeq 20$
at $\tau/t = 0$, $0.35$ and $0.65$. The error bars are small, and the
traveling ramp structure is very well pronounced.
Figure~\ref{high_com} shows, at $\tau=0.5 t$,  that the
theoretically-predicted asymptotic optimal path \cite{Janas2016}, described by
Eqs.~(\ref{ramp1})-(\ref{ramp3}), is in good agreement with  the
simulated ramp profile. The observed discrepancies ($\lesssim 5\%$ in
terms of $H$) can be attributed to effects of finite $1/H$ and $H/L$,
see Eq.~(\ref{mapconditions}).

\begin{figure} [ht]
\includegraphics[width=0.36\textwidth,clip=]{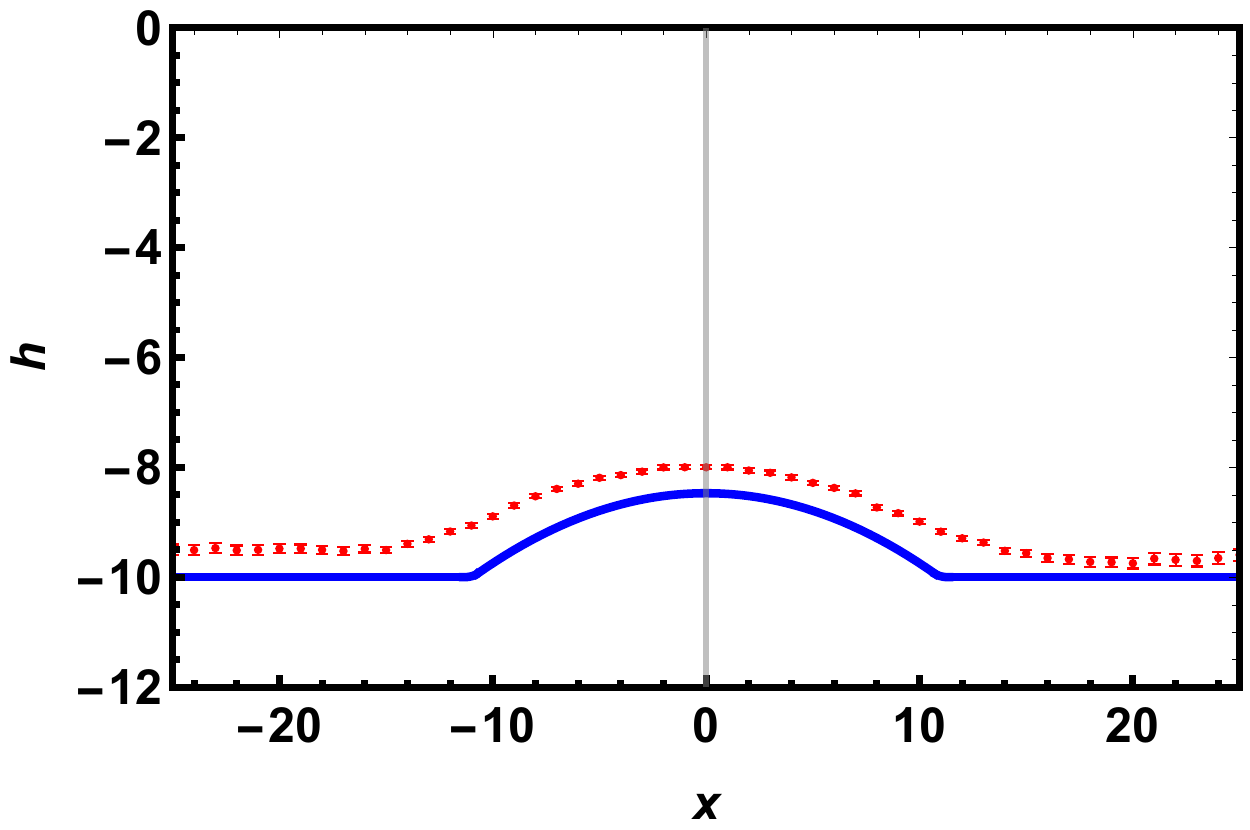}
\includegraphics[width=0.36\textwidth,clip=]{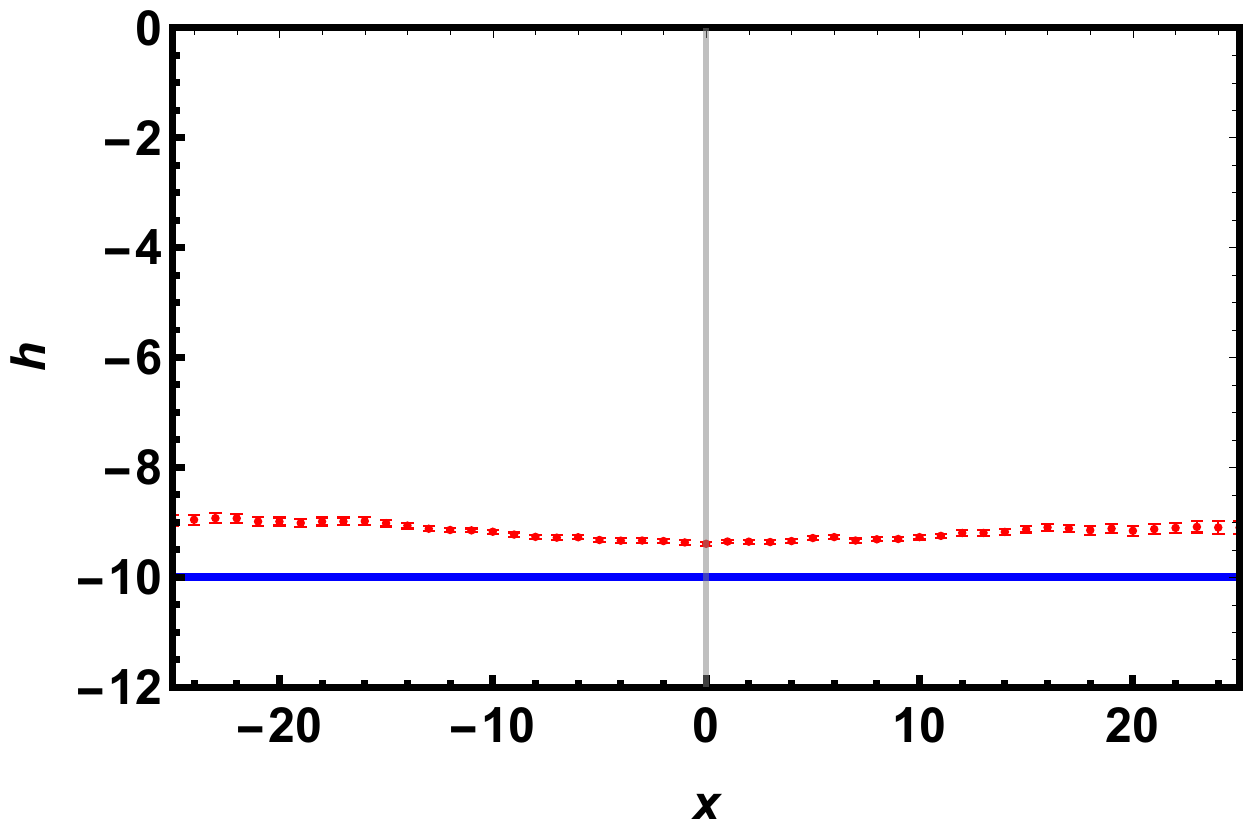}
\includegraphics[width=0.36\textwidth,clip=]{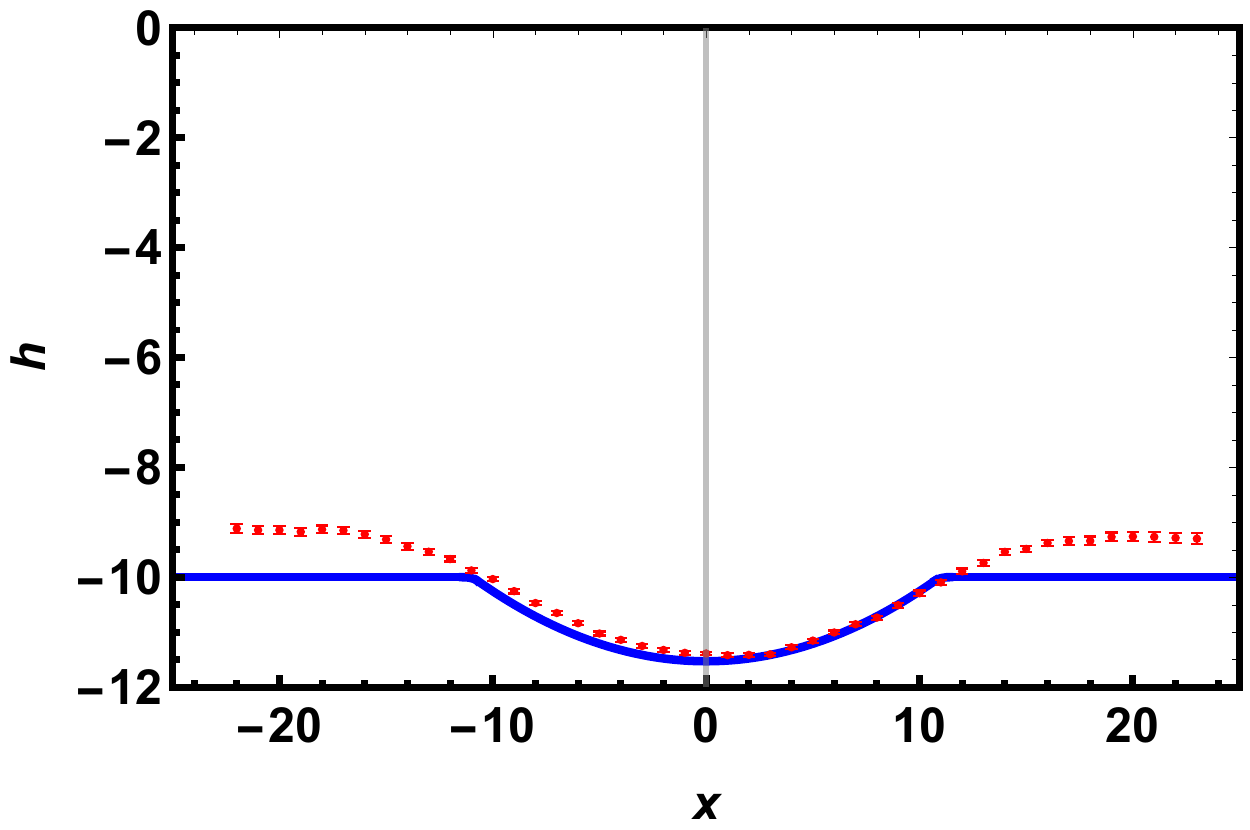}
\caption{A comparison of the simulated average height profile conditioned on $H\simeq
-20$ (the red points with error bars) with the asymptotic solution~(\ref{HD1})-(\ref{HD3}) for the optimal path $h(x,\tau)$ at rescaled times $\tau/t=0.35$ (top), $0.5$ (middle)
and $0.65$ (bottom) for $T=8$ (which corresponds to a short time) and $L=64$.}
\label{low}
\end{figure}

\subsubsection{Large negative $H$}

Figure~\ref{low} shows the average over 60 simulated $h$-profiles for $H\simeq -20$ (the
actual values $H\in[-20.2,-20]$). Realizations exhibiting such values of $H$ appear with natural probabilities about $10^{-450}$ or smaller \cite{HKLD}.
Also shown is the theoretically-predicted asymptotic optimal path for $H=-20$, see Eqs.~(\ref{HD1})-(\ref{HD3}). The mirror symmetry is manifestly preserved
here. Again, the discrepancy between the asymptotic theory and simulations is of order of 5\% in terms of $|H|$, and it can be attributed to finite  $|H|/L$ effects. Indeed, we also simulated the case of $H\simeq -40$ (not shown here) and found that the discrepancy between the theory (which assumes $|H|\ll L$)  and the simulations becomes quite large.

\section{Long times} 
\label{longtime}

For arbitrary values of $H$, the  OFM provides an asymptotically exact large-deviation function of the KPZ height $S(H)$ see Eqs. (\ref{actiondgen}) and (\ref{LDF}), in the limit of $t\to 0$, or $\epsilon \to 0$. However, it was argued in Ref. \cite{MKV}, that the OFM should remain accurate in its description of (sufficiently far) distribution \emph{tails} at \emph{any} time $t$, or at any $\epsilon$. Since then it has been firmly established  -- by different methods which include rigorous mathematical proofs -- that, for a broad class of initial conditions, the distribution tails, as
predicted by the OFM, are indeed  observed at all times \cite{LDMS2016,SMP,Co18,Ts18,KLDP,CorwinGhosal,D20,FerrariVeto2021,GhosalLin}. This suggests (but does not guarantee) that the dominant role of the optimal paths
  and their shape can be described by the OFM at long times as well,
and we address this issue shortly.

Furthermore, there is an interesting difference in the structure of the lower and upper tails of $\mathcal{P}(H,t)$ at long time. This difference has been thoroughly studied (including rigorous proofs) for a class of
deterministic initial conditions and, first of all, for the droplet initial condition \cite{LDMS2016,SMP,Co18,Ts18,KLDP,CorwinGhosal,D20,GhosalLin}.  The lower tail has a double structure. It consists of two distinct tails: the near and the far.  The near tail, which is located at $1\ll t^{1/3}\ll |H|\ll t$,  scales as
$-\ln \mathcal{P} \sim |H|^3/t$. The far tail, located at $1\ll t \ll |H|$,  exhibits the OFM scaling (\ref{Slowtail}): $-\ln \mathcal{P} \sim H^{5/2}/\sqrt{t}$. (For brevity we dropped here the factors $\lambda, \nu$ and $D$.) An optimal path, predicted by the OFM, was indeed observed in our previous simulations of the far tail for the droplet initial condition \cite{HMS1}, but no observations  of the interface profiles corresponding to the near tail have been performed yet, for any initial condition.

The situation with the upper tail at long times is even less clear. This tail also has a near and a far regions, located at $1\ll t^{1/3}\ll |H|\ll t$ and  $t\ll |H|$, respectively. However, the form of the tail in these two regions turns out to be the same up to subleading corrections, and it is described by Eq.~(\ref{Shightail}). For stationary initial condition this was shown in Ref. \cite{Janas2016} by extracting the corresponding asymptotic of the Baik-Rains distribution \cite{BR} which, according to Refs. \cite{IS,Borodinetal}, describes typical fluctuations of the height at long times  \cite{BRhistory}.  The same situation, without exceptions, occurs for all other initial
conditions considered so far (for the flat initial condition this has been known since Refs. \cite{KK2007,KK2008}). The coincidence of the near and far tails raises the question of whether a well-defined ``dominating path" of the interface exists
only in the far region of the upper tail, $H\gg t$, or in the near region too.

To shed light on the latter issue, we extended our numerical simulations of the interface dynamics to a lower temperature, $T=2$, keeping $L=64$. For these parameters we obtain $\epsilon=32$ which already corresponds to the long-time limit. We conducted two series of simulations: (a) conditioned on $H\simeq 20$, and (b) conditioned on $H\simeq 5$. The value of $H\simeq 20$  is not far from the regime $H> t$, and here one can expect to see an optimal path similar to that predicted by the OFM. The value of $H\simeq 5$ is already in the non-OFM regime of $t^{1/3}<H<t$.

\begin{figure} [ht]
\includegraphics[width=0.36\textwidth,clip=]{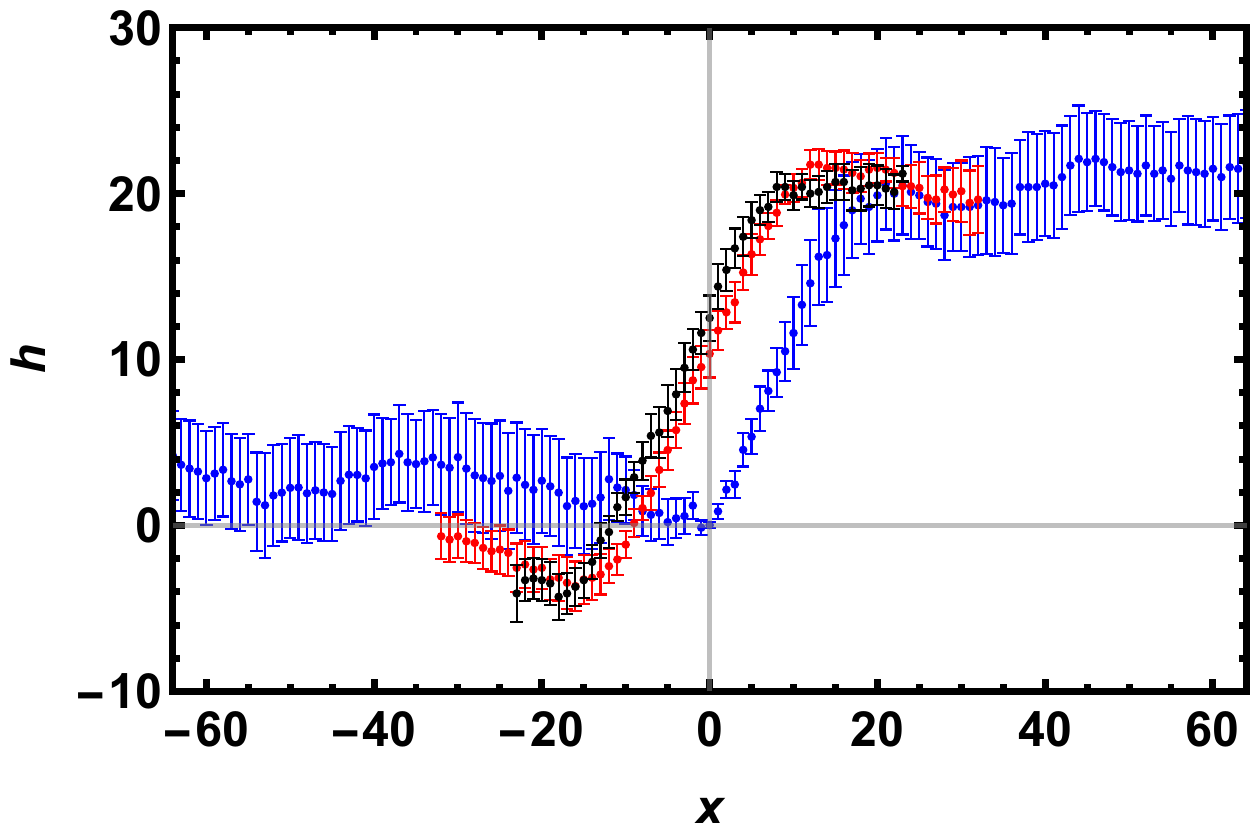}
\includegraphics[width=0.36\textwidth,clip=]{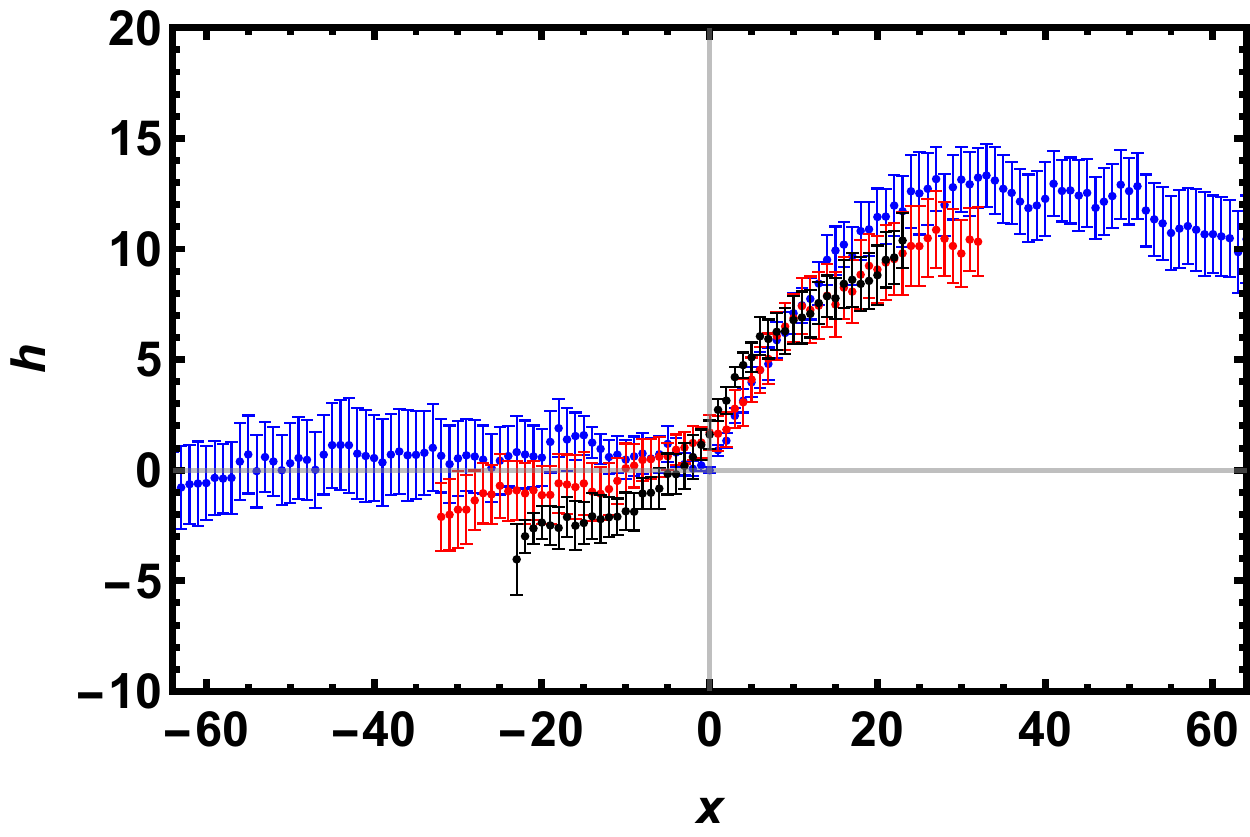}
\caption{The average simulated height profiles $h(x,\tau)$ conditioned on $H\simeq 20$ (top) and $H\simeq 5$ (bottom) at rescaled times $\tau/t=0$ (blue), 0.5 (red) and 0.65 (black). There were 11 measured profiles for $H\simeq 20$ and 8 profiles for $H\simeq 5$.  The actual values of $H$ are
  $H\in [18.81, 20.07]$ (top) and  $H\in [4.41, 5.18]$ (bottom).  5 of the height profiles, both for $H\simeq 20$ and $H\simeq 5$, were flipped around $x=0$.  The error bars show the standard deviation. The simulation parameters $T=2$ and $L=64$ correspond to a long-time regime.}
\label{longtimefig}
\end{figure}

\begin{figure} [ht]
\includegraphics[width=0.36\textwidth,clip=]{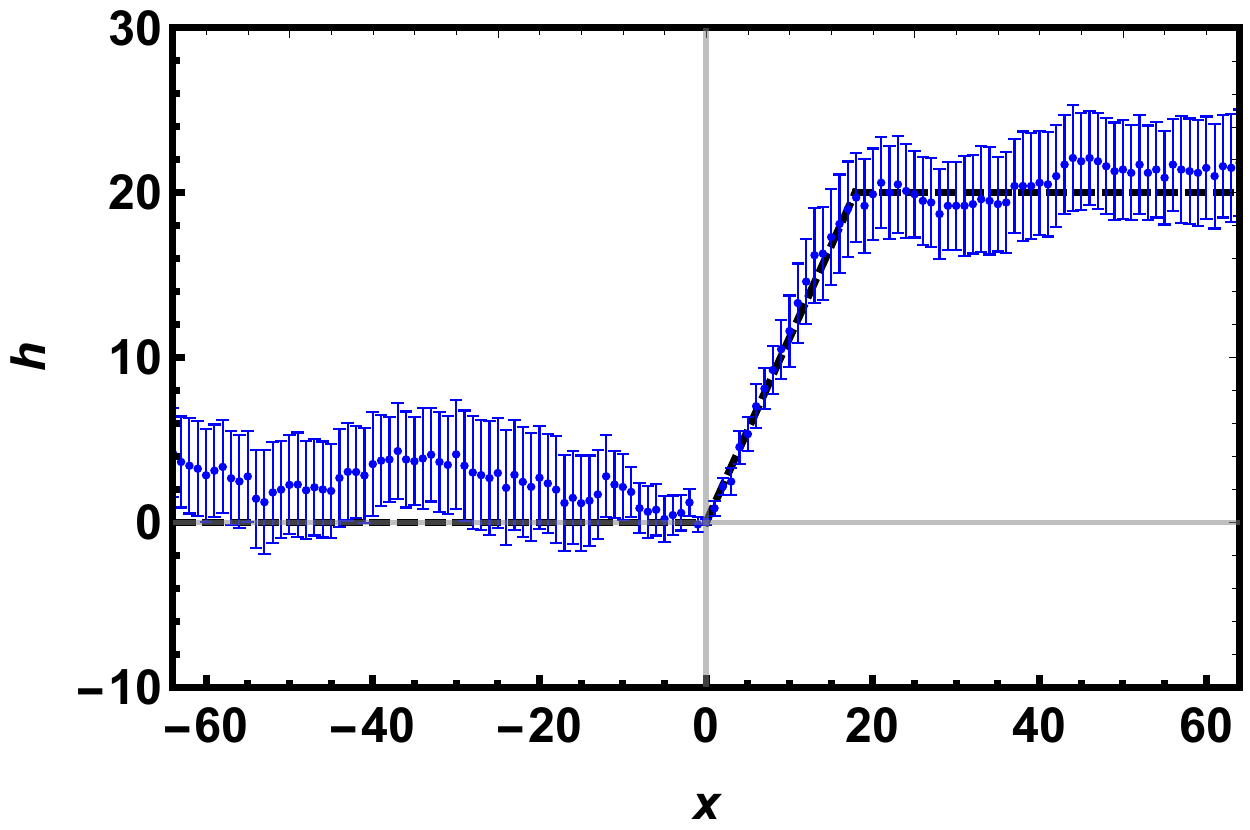}
\includegraphics[width=0.36\textwidth,clip=]{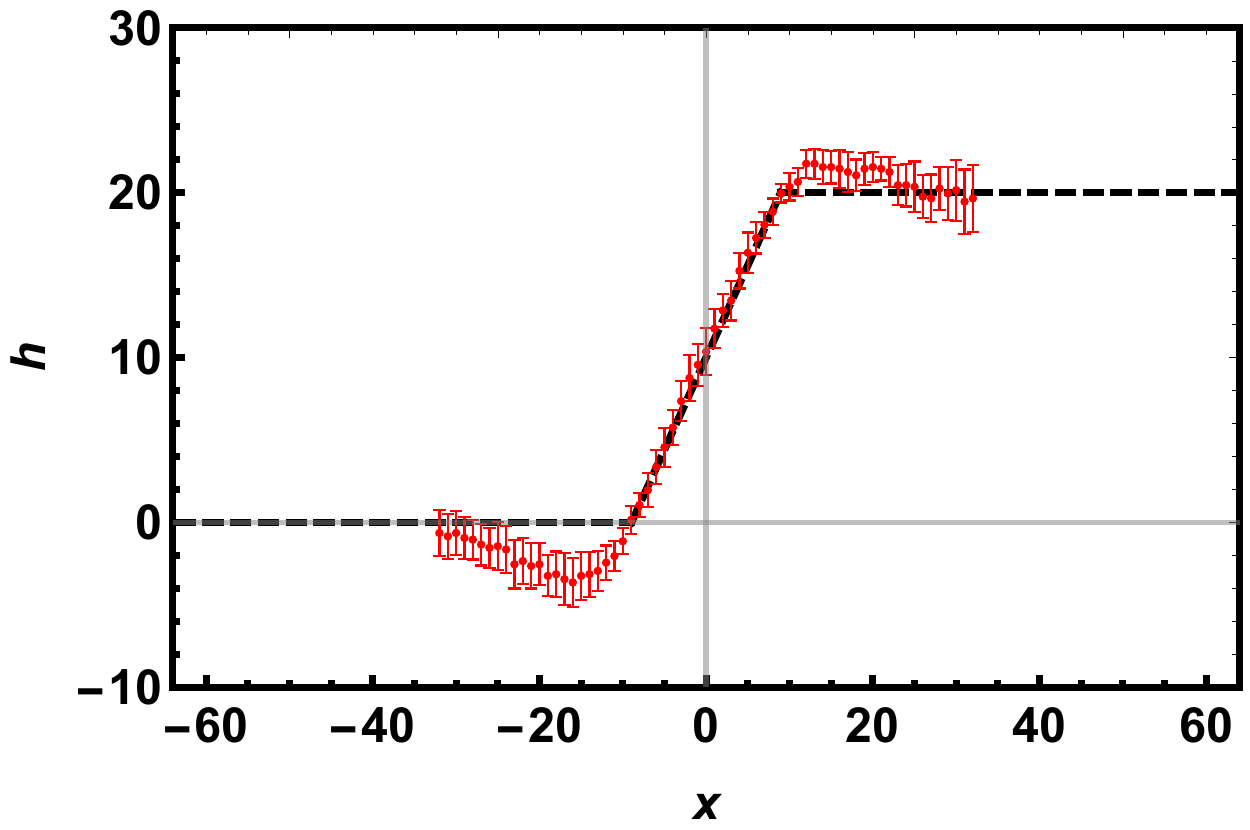}
\caption{A comparison of the simulated average height profiles conditioned on $H\simeq 20$ [also shown in the upper panel of Fig. \ref{longtimefig}]  with the asymptotic
solution (\ref{ramp1})-(\ref{ramp3}) for the optimal path \cite{Janas2016}, shown by the dashed lines, at times $\tau/t=0$ (top) and $0.5$ (bottom). The simulation parameters $T=2$ and $L=64$ correspond to a long-time regime. See Fig. \ref{longtimefig} for other details.}
\label{longtimefigcomp}
\end{figure}

Our simulation results for $T=2$ are depicted in Figs. \ref{longtimefig} and \ref{longtimefigcomp}. For $H\simeq 20$ [Fig. \ref{longtimefig} (top)]  we indeed observed two optimal paths which agree fairly well with the symmetry-broken solutions -- the two traveling ramps -- predicted by the OFM, see Fig. \ref{longtimefigcomp}. There are, however, some noticeable differences, clearly seen in the bottom panel of Fig. \ref{longtimefigcomp}. The larger noise in the profiles in the left and right plateau regions, in comparison with the simulations for $T=8$, can be explained by the larger (by a factor of 4) diffusion constant of the Brownian initial condition in this case.

\begin{figure} [ht]
\includegraphics[width=0.36\textwidth,clip=]{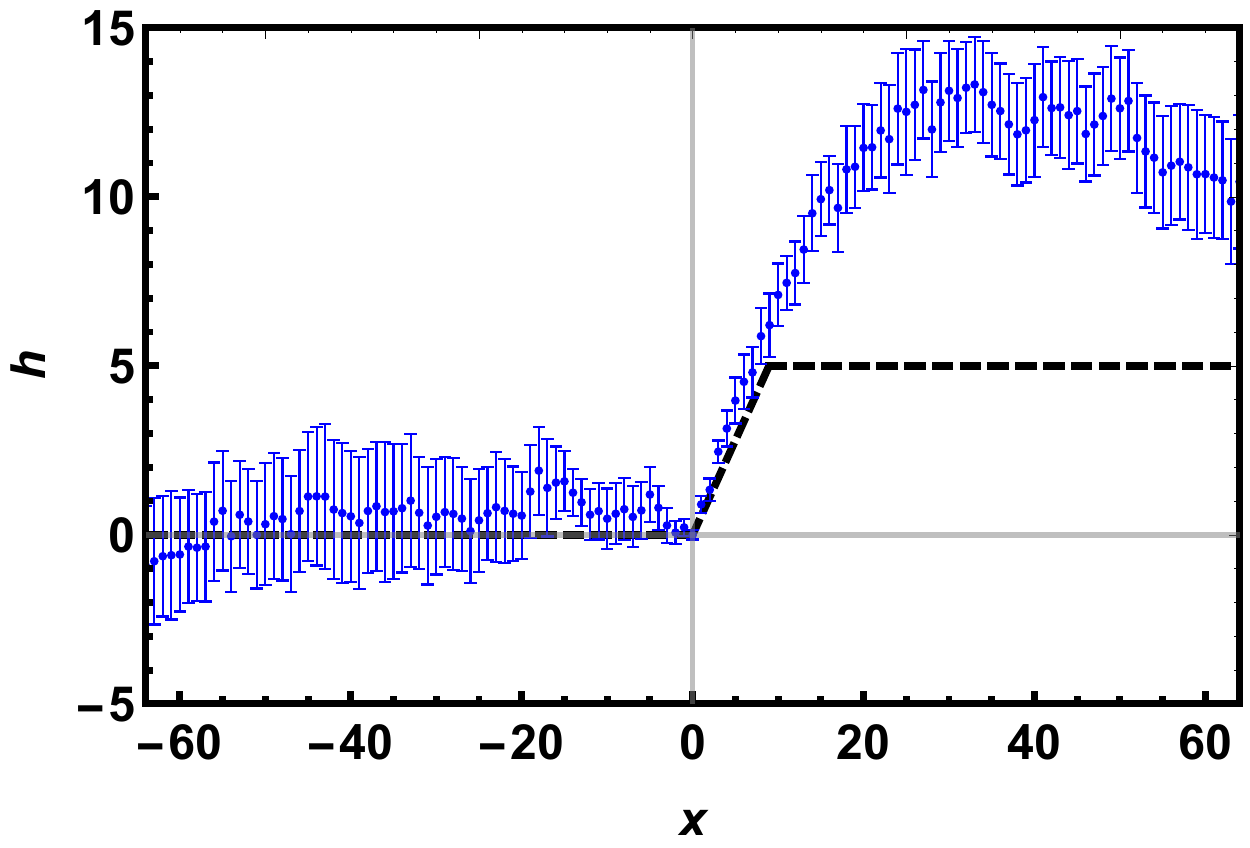}
\caption{The simulated average height profile conditioned on $H\simeq 5$ at long times (also shown in the upper panel of Fig. \ref{longtimefig}) and the asymptotic
solution (\ref{ramp1})-(\ref{ramp3}) for the optimal path \cite{Janas2016} (the dashed line), at time $\tau=0$.  The simulation parameters $T=2$ and $L=64$ correspond to a long-time regime. There is a clear quantitative disagreement between the OFM prediction and simulation results, but the mere presence of two well defined, mutually symmetric dominating paths of the interface (where one of them represents a mirror image of the other) is noteworthy.}
\label{longtimefigcomp2}
\end{figure}

Surprisingly, at $H\simeq 5$ (the bottom panel of Fig. \ref{longtimefig}), we still observed two well-defined paths with a broken mirror symmetry. Qualitatively, these ``dominating paths" $h(x,\tau)$ resemble the optimal paths predicted by the OFM, but there is no quantitative agreement anymore, see Fig. \ref{longtimefigcomp2}.  Still, the mere existence of two symmetry-broken dominating path of the conditioned process in this non-OFM regime
is remarkable. Moreover, each of these two paths is again a mirror image of the other.

\section{Summary and Discussion}
\label{summary}

Optimal paths provide an instructive and fascinating characterization of a whole class of large deviations in
non-equilibrium stochastic systems. By combining a mapping between the KPZ interface and the directed polymer in a random potential at high temperature with a large-deviation Monte Carlo
sampling algorithm, we were able
to observe the optimal paths of the KPZ interface which determine each of the two tails of
the one-point height distribution $\mathcal{P}(H,t)$ at short times,
and extended the notion of the dominating
path to long times. Our algorithm allowed us to probe the optimal
paths which are responsible for extremely unlikely events, with probability densities down to $10^{-500}$.

In the short-time regime we clearly observed mirror-symmetry-broken  ``traveling-ramp" optimal interface height profiles for the upper tail, and mirror-symmetric paths for the lower tail, in good quantitative agreement with analytical predictions from the OFM.

In the long-time regime we identified two different regions in the upper tail of $\mathcal{P}(H,t)$. In the far region (very large $H$) we observed an optimal path in fair agreement with the predictions of the OFM, although some noticeable differences appear.  In the near region (moderately large $H$) the OFM is invalid. Still, we clearly observed two  well-defined dominating paths of the interface.  Each of them violates the mirror symmetry and represents a mirror image of the other. The presence of dominating paths of the interface hints at a possible existence of a \emph{renormalized }optimal fluctuation theory for the near tails in the late-time regime of the KPZ equation.

Simulation-wise our work, alongside with Ref. \cite{HMS1},  shows that, by using large-deviation approaches, one can not only reliably determine extremely small probabilities but also sample complete realizations from the extreme tails of corresponding distributions. This gives a valuable insight into the causes of extreme behavior of stochastic systems far from equilibrium. In the present case this is especially clear in the short-time regime, because here the OFM predicts an intimate connection between extremely small probabilities of observing very large $|H|$ and extremely rare interface configuration histories that
are responsible for them. But the possibility of determining causal relations
certainly also exists when no theoretical predictions are available, as was already shown earlier,
\textit{e.g.}, in large-deviation studies of stability conditions for energy grids \cite{power_flow2015,stability2019}. Furthermore, even if no clear causal relations emerge, large-deviation approaches allow for a much enlarged view into correlations in complex systems.

\vspace{0.5 cm}

\section*{Acknowledgments}
The simulations were performed in Oldenburg on the HPC cluster CARL which
is funded by the DFG
through its Major Research Instrumentation Programme (INST 184/157-1 FUGG) and the Ministry of
Science and Culture (MWK) of the Lower Saxony State. The research of B.M.
was supported by the Israel Science Foundation (grant No. 1499/20). The research of P.S. is supported by the project ``High Field Initiative" (CZ.02.1.01/0.0/0.0/15\_003/0000449) of the European Regional Development Fund.


\begin{thebibliography}{99}
\bibitem{pinning} Without loss of generality, we can pin the Brownian interface at the origin so that $h(x=0,\tau=0)=0$.
\end{thebibliography}
\end{document}